\documentstyle[11pt,aaspp4,tighten]{article}
\newcommand{\href}[2]{#2 (\texttt{#1})}
\def\rosat{{\sl ROSAT~}}
\def\asca{{\sl ASCA~}}

\slugcomment{}
\begin{document}

\title{\bf Under the Shadow of the Magellanic Bridge: A Measurement
of the Extragalactic Background at $\sim 0.7$~keV}

\author{Q. Daniel Wang}
\affil{Dearborn Observatory, Northwestern University}
\affil{ 2131 Sheridan Road, Evanston,~IL 60208-2900}
\affil{Electronic mail: wqd@nwu.edu}
\affil{and}
\author{Taisheng Ye}
\affil{Australia Telescope National Facility}
\affil{P.O. Box 76, Epping, NSW 2121}
\affil{Electronic mail: tye@atnf.csiro.au}

\begin{abstract}

	We measured the extragalactic 0.7~keV X-ray background by observing
the X-ray shadow of a neutral gas cloud in the Magellanic Bridge region. 
Two \rosat PSPC observations of total 104~ks were complemented by a
detailed  \ion{H}{1} mapping of the cloud with 
both the Parkes 64~m telescope and the Australia Telescope Compact Array.
From the detected anti-correlation between the observed background intensity
and the \ion{H}{1} column density of the cloud, 
we derived the unabsorbed extragalactic background intensity as $\sim 28
{\rm~keV~s^{-1}~cm^{-2}~keV^{-1}} {\rm~sr^{-1}}$
at $\sim 0.7$~keV. The 95\% confidence lower limit $18
{\rm~keV~s^{-1}~cm^{-2}~keV^{-1}} {\rm~sr^{-1}}$ is greater than the expected
 point-like source contribution $\lesssim 14 
{\rm~keV~s^{-1}~cm^{-2}~keV^{-1}} {\rm~sr^{-1}}$, constrained by 
the mean source spectrum together with the total background intensity
in the 1-2~keV band. A significant fraction of the 0.7~keV 
background likely arises in a diffuse hot intergalactic medium 
of a few million degrees, as has been predicted in 
hydrodynamic simulations of cosmological structure formation. 

\end{abstract}


\section {Introduction}

	The extragalactic background at $\sim 0.7$~keV
is of special cosmological interest, because it may contain unique
information on the large-scale structure of the universe.
Hydrodynamic simulations of the structure formation 
predict that a considerable fraction, probably most, of the baryonic 
matter in the universe is in a diffuse intergalactic medium (IGM) within 
a temperature range of $10^5$-$10^7$~K 
(Ostriker \& Cen 1996; OC96 hereafter). This ``not-so-hot'' IGM,
not precluded by the COBE spectral distortion limit on the cosmic 
microwave background (Mather et al. 1994), may contribute significantly 
to the soft ($\lesssim 1$~keV) X-ray background. Most interestingly, 
various emission lines of heavy elements (O, Ne, and Fe) may collectively form 
a distinct spectral bump at $\sim 0.7$~keV (Cen et al. 1995; CKOR95
hereafter). This bump is an observationally testable feature, and 
has the potential as a diagnostic of the physical and chemical properties 
of the IGM. 

	The directly observed 0.7~keV background is, however, a composite
of various components, both Galactic and extragalactic. The 
relatively well-studied component is from discrete sources, most 
($\gtrsim 90\%$)  of which appear to be AGNs (e.g., Shanks et al. 1991). 
Based on the \rosat deep and medium sensitivity surveys, 
Hasinger et al. (1993; H93 hereafter) have determined the source
flux distribution down to a 0.5-2~keV flux limit of $\sim 2.5 \times 10^{-15} 
{\rm~ergs~s^{-1}~cm^{-2}}$. The distribution can be described by 
a broken power law, which changes its slope at a source flux of 
$\sim 2.5 \times 10^{-14} {\rm~ergs~s^{-1}~cm^{-2}}$.
The average source spectrum in the range of 0.1-2~keV can be 
characterized by a power law of energy slope $\alpha = 0.96\pm0.11$, but
the spectrum becomes harder with the decreasing source flux
(see also Wang \& McCray 1994; Vikhlinin et al. 1995). 
Spatial fluctuation analysis of X-ray data further
suggests that point-like sources account for $\gtrsim 75\%$ of the background
intensity in the 1-2~keV band (H93). In the lower energy bands, however,
discrete sources appear to contribute a significantly lower fraction.
Based on auto-correlation and spectral analyses of two \rosat observations, 
Wang \& McCray (1994) conclude that point-like sources account for 
$\lesssim 60\%$ of the 0.7~keV  background below a source detection limit
of $1 \times 10^{-14} {\rm~ergs~s^{-1}~cm^{-2}}$, and suggest the presence of 
a diffuse ``hard'' component, in addition to the well-known $10^6$~K Local 
Bubble contribution that is important only at energies $\lesssim 0.3$~keV 
(Snowden et al. 1990). The nature of this diffuse hard component, 
however, remains unclear.
	
	The Galactic contribution must be important at least in some 
directions. The \rosat all sky survey (Snowden et al. 1995) shows a
definite intensity enhancement in the 3/4~keV band 
(approximately the Wisconsin M band) towards the inner portion of the Milky Way
($\lesssim 90^\circ$ from the Galactic center). The Galactic contribution 
must be dominant at the Galactic plane, which is nearly opaque to the 
extragalactic M-band radiation. Although its exact origin is not 
yet clear, the contribution appears to be chiefly diffuse (Wang 1992).

	To separate the Galactic and extragalactic components of
the M-band background, we carried out an observing program to measure
the background shadow produced by the X-ray absorption of a cloud in the 
bridge region between the Large and Small Magellanic Clouds (LMC and SMC;
McGee \& Newton 1986). This cloud, about $4^\circ$ away from the SMC, is 
clearly a product of the tidal interaction between the two 
\ion{H}{1} galaxies. Besides the \ion{H}{1} cloud, the field 
appears to be a fair sample of the typical sky, free from X-ray emission 
associated with the LMC and SMC. Furthermore, at a distance 65~kpc,
assumed to be the same as the SMC, the cloud should shadow the 
true extragalactic background. Most importantly,
the peak column density of the cloud, $\sim 3 \times 10^{21} {\rm~cm^{-2}}$, 
is only about 50\% transparent to X-rays at $\sim 0.7$~keV . Therefore, 
we can infer the extragalactic component from the X-ray absorption by the 
cloud.

\section {Observations and Data Reduction}
\subsection {X-ray} 

	Two of our proposed three observations were carried out
with the {\sl ROSAT} Position Sensitive Proportional Counter
(PSPC --- Pfeffermann et al. 1987; Tr\"umper 1983): The on-cloud 
observation (\rosat Sequence No. rp900250) with 
an accumulated exposure of 91~ks was centered at
$2^h 17^m.2, -74^\circ 06^\prime$ (R.A., Dec. --- J2000). The only off-cloud 
observation of 13~ks (considerably less than desirable) was centered at
$2^h 22^m, -75^\circ 33^\prime$.
The fields of view of the two observations, $2^\circ$ diameter each,
were set to overlap. The overlapping region allowed us to remove differential
contamination of non-cosmic X-rays between the two observations.

	Standard data reduction for diffuse study included
removal of times of high solar X-ray contamination, estimation of residual 
non-cosmic X-ray background, and flat-fielding
(Wang \& Yu 1995; Snowden et al. 1994a). The observation 
rp900250 was taken in two different observing 
semesters: March 16 to April 15, 1993 and October 1 to November 1, 1993. These
two data sets were first analyzed separately, and later aligned and co-added 
by matching discrete source positions (a correction of 
$\sim 10^{\prime\prime}$). 

	Count, non-cosmic X-ray background, 
and exposure images were produced in individual \rosat PSPC
bands (i.e., R1-R2, R4-R7; Snowden et al. 1994a). The R1 and R2 bands were
sensitive to photons of energies $\lesssim 0.284$~keV, the Carbon-absorbing 
edge of the detector window. Photons observed in these two bands are 
dominantly 
Galactic; the sightline Galactic absorption $\sim 4.3 \times 10^{20} 
{\rm~cm^{-2}}$ corresponds to a mean photon transmission $\lesssim 5\%$.  
The R4 band, whose response covers an energy range of
0.44--1.0~keV and peaks at 0.7~keV (approximately the Wisconsin M1 band), 
is optimal for detecting the extragalactic 0.7~keV background absorption 
by the bridge cloud. 
The rest of the bands response more favorably to photons at higher energies
(0.6--2~keV; the energy responses of the bands are partially overlapped
due to the limited spectral resolution of the PSPC).
The transmission increases steeply
with the increasing photon energy (Morrison \& McCammon 1983). 
In the R5 band, for example, the transmission is about 50\% greater than 
in the R4 band.

\subsection {Radio}

We mapped out the \ion{H}{1}
column density distribution in the \rosat fields using the Australia 
Telescope Compact Array (ATCA) and Parkes 64~m telescope.
The Parkes observations were made during August 29 and October 23-24, 1993.
For the on-cloud field, the observations covered an area of 
$50^{\prime}$ radius with a sample interval of 1$^m$ in R.A. and 
4$^{\prime}$ in Dec. 
For the off-cloud area, we observed 5 positions with separation 
of 5$^m$ in R.A. and 20$^{\prime}$ in Dec. The integration time at each 
position was 30~s, and a total of 561 pointings was obtained. 

	While the Parkes mapping was used to determine the total
\ion{H}{1} distribution of large scale structure, the ATCA observations
toward the cloud were used to resolve fine features. The observations
were made on July 28, 1993 with the configuration 750D
(the shortest baseline is 30~m). Centered at 1420 MHz, the 8~MHz bandwidth
was divided into 1024~channels. PKS 1934-638
(adopted flux density of 16.1~Jy) was observed for the absolute flux 
density calibration, while PKS 0252-710, observed 
for 5 minutes at every 30 minute interval, 
was used for phase and spectral bandpass calibrations. 

	These radio observations were processed with the AIPS 
software package. The Galactic \ion{H}{1} component 
(velocity $\lesssim 160 {\rm~km~s^{-1}}$) was detected only in the 
Parkes observations, indicating that it is very smooth and has
few small-scale features. We adopted an average Galactic column
density of $4.3 \times 10^{20} {\rm~cm^{-2}}$ 
with an uncertainty of $\sim 10\%$. 
In contrast, structures with various angular scale 
sizes are present in the bridge component, which was integrated 
over the velocity range of $ 160 - 220 {\rm~km~s^{-1}}$.

	To obtain both the absolute column density and its
fine structures, we combined the Parkes and ATCA data
to form a single \ion{H}{1} map of the cloud (Fig.~1), using the 
method of adding single-dish data into an aperture synthesis data as
described by Ye, Turtle and Kennicutt (1991). The resolution of this map
is $50^{\prime\prime}$ which is comparable to that of 
the X-ray data. Smoothing the map back 
to the resolution of the Parkes showed no distinct difference from the 
one produced by the Parkes data alone. 

\begin{figure} 
\figurenum{1}
\plotone{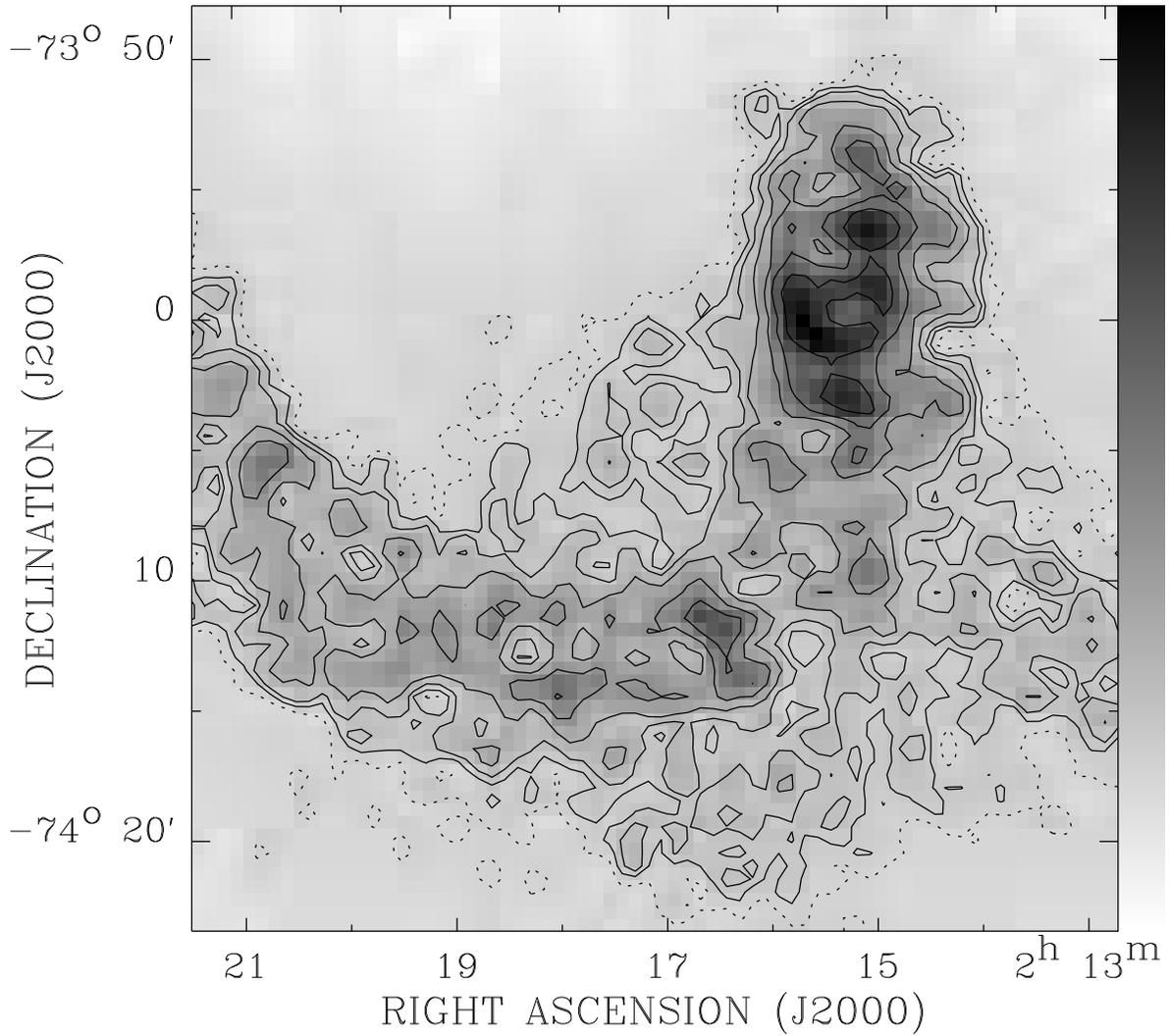}
\caption{(URL: 
http://www.astro.nwu.edu/astro/wqd/bridge/fig1c.gif) 
Atomic hydrogen column density map of 
our target cloud in the Magellanic Bridge region. The contours are at 11, 
12, 13, 15, 18, 22, and $27 \times 10^{20}$ $ {\rm~cm^{-2}}$.
\label{fig1}}
\end{figure}
\section {Analyses and Results}

As shown in Fig.~2,  discrete sources cause strong small-scale fluctuation 
in the X-ray intensity map. Thus, we must remove them before performing an
X-ray/\ion{H}{1} anti-correlation analysis. We searched for sources in each 
of the two PSPC observations. We first used the so-called local detection 
algorithm (Hamilton, Helfand, \& Wu 1992), and then used the standard map 
detection for more sensitive searches. We constructed a background 
map by excising detected sources and by smoothing with a median
filter (Wang \& Yu 1995). We adopted two source detection apertures:
50\% and 85\% flux-encircled radii of the instrument point spread function
(PSF; Hasinger et al. 1994).  We applied the maximum likelihood fitting
to obtain accurate positions of those sources 
within $18^\prime$ on-axis of each observation. Outside this 
off-axis radius, the 
quality of source detection and analysis was compromised
by both the degraded PSF and the shadow of the detector's window 
supporting structure. Accordingly, our later anti-correlation analysis 
is limited to the data within the off-axis radius. Finally, 
we merged sources within the 50\% flux-encircled radius. We repeated
the map detection until no new source was found.

\begin{figure} 
\figurenum{2}
\plotone{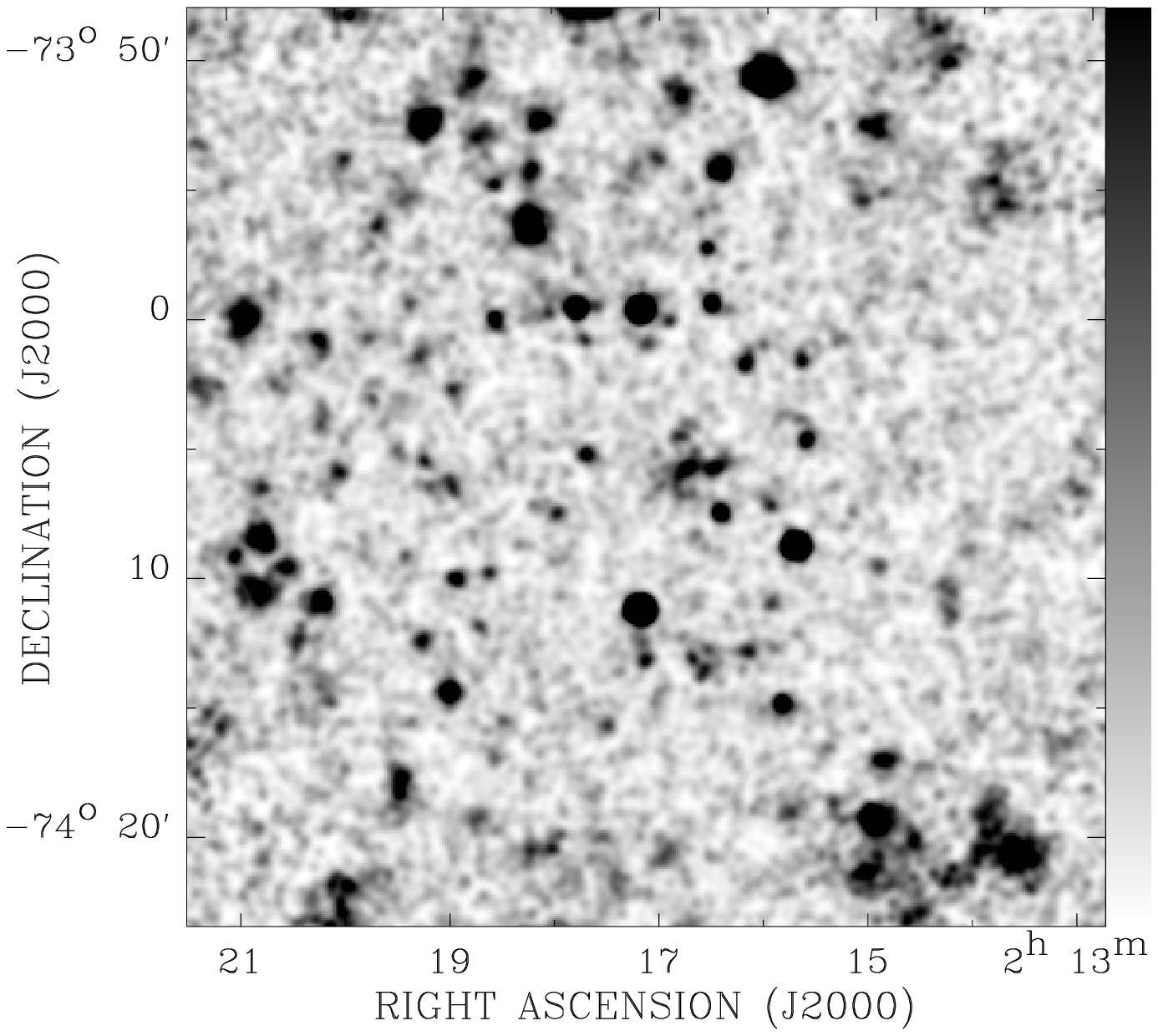}
\caption{(URL: 
http://www.astro.nwu.edu/astro/wqd/bridge/fig2c.gif)
Background-subtracted and exposure
corrected X-ray image of the on-cloud field. The image
is constructed in the broad 0.5-2~keV band (R4-7) and only the central
portion of the on-cloud observation is shown.
The image, smoothed with a Gaussian of size 25$^{\prime\prime}$, 
is scaled in the range of 0-2 $\times 10^{-3} $ $
{\rm~counts~s^{-1}~arcmin^{-2}}$ (yellow is the brightest).
\label{fig2}}
\end{figure}

	We removed sources brighter than an unabsorbed flux of 
$7 \times 10^{-15} {\rm~ergs~s^{-1}~cm^{-2}}$ (0.5-2~keV), 
which corresponds to a source detection 
signal-to-noise ratio of 3 at the $18^\prime$ off-axis of the on-cloud 
observation. In converting from a PSPC count rate to its 
unabsorbed flux, we  assumed a power law spectrum of $\alpha =1$, absorbed 
by \ion{H}{1} in the source's direction. Actually, we excluded only 
data within the $90\%$ flux-encircled radius of each source; the 
remaining 10\% of the source 
flux, contained mainly in the broad PSF wing due to irregular 
scattering of X-rays by the imperfect telescope mirror, cannot be effectively 
subtracted on a pixel-to-pixel base. Such residual source flux, 
together with sources below our source removal threshold, 
causes some small-scale enhancements in Fig.~3. 
In the off-cloud observation, we removed sources down to a threshold 
of $2 \times 10^{-14} {\rm~ergs~s^{-1}~cm^{-2}}$. To calculate 
intensities in the off-cloud observation at the same source removal 
threshold as in the on-cloud observation, we further subtracted 90\% of
the source contribution between the two thresholds, which we estimated 
with the Hasinger et al source flux distribution, an approach 
similar to that used in Cui et al. (1996).

\begin{figure} 
\figurenum{3}
\plotone{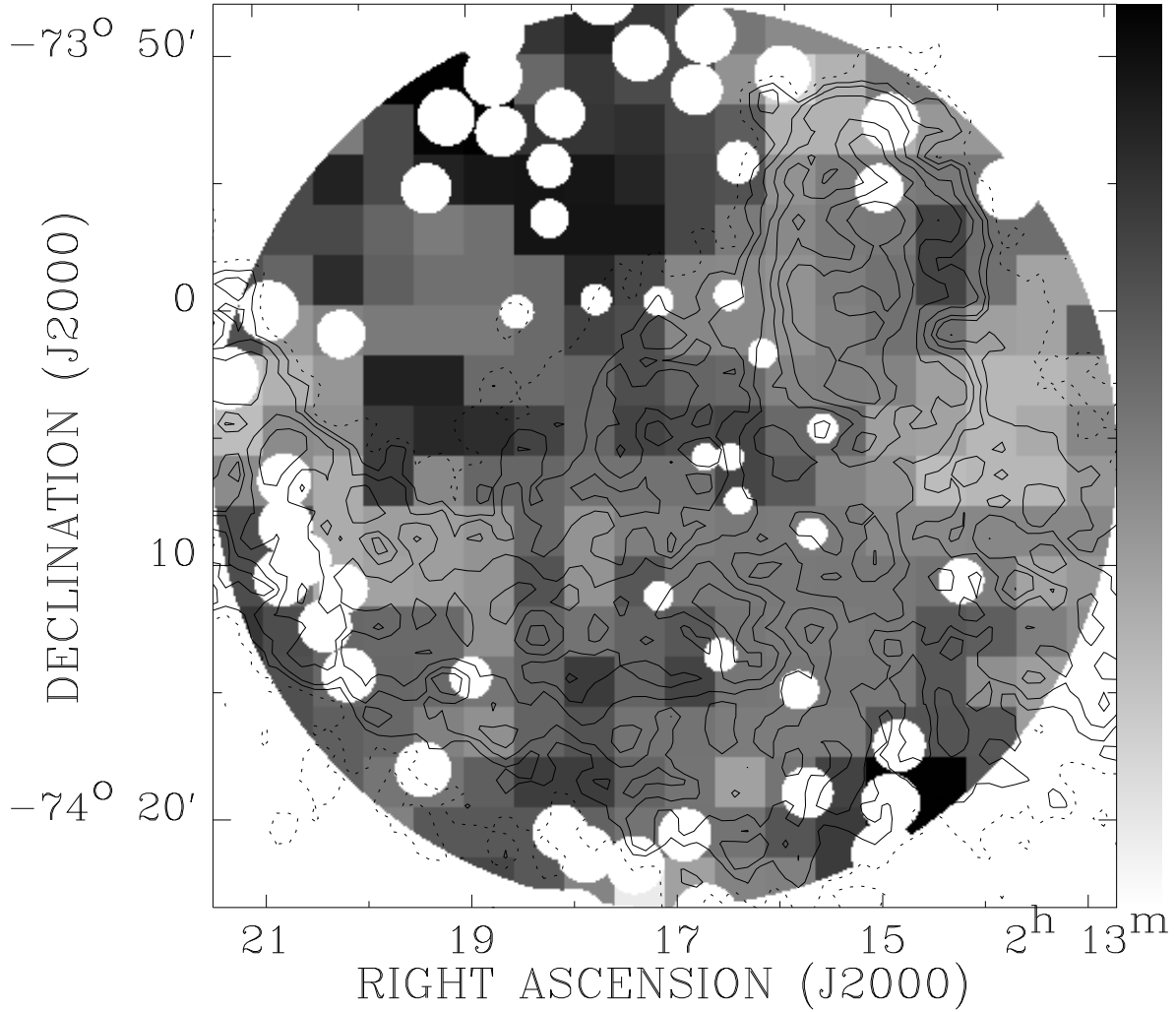}
\caption{(URL: 
http://www.astro.nwu.edu/astro/wqd/bridge/fig3c.gif)
Source-removed image of the cloud in the PSPC R4 band, 
overlaid by atomic hydrogen column density contours. Small circles 
are the regions within which data have been excluded for individual sources.
The image, smoothed with a median filter (Wang \& Yu 1995), is scaled in 
the X-ray intensity range of 2-8 
$\times 10^{-5} $ ${\rm~counts~s^{-1}~arcmin^{-2}}$. The 
contours are the same as in Fig.~1. 
\label{fig3}}
\end{figure}

	To facilitate a least-$\chi^2$ analysis, we summed the on-cloud 
data in each band into 100 \ion{H}{1} column density bins of 
about equal area coverage (e.g., average 65.7 counts per bin in the R4 band), 
and the off-cloud data into a single bin. We 
fitted the data with a two-component model 
(Fig.~4), defined as
  $$I_t^i= I_g + I_e T^i,    $$
where $I_g$ and $I_e$ are
the Galactic and extragalactic intensities to be determined; 
$I_t^i$ is the total expected X-ray intensity within an \ion{H}{1} 
bin, denoted by superscript $i$, and $T^i$ is the mean X-ray transmission 
in the bin. The transmission accounts for the neutral gas absorption 
in both the bridge and the Galaxy.  Elements contributing 
to the absorption are primarily H, He, O, and Ne
(Morrison \& McCammon 1983).  We adopted a metallicity 
of 18\% solar for the cloud, same as that of the SMC (Russell
\& Dopita 1992), and 50\% 
solar for Galactic \ion{H}{1}  (Meyer et al. 1995).

\begin{figure} 
\figurenum{4}
\plotone{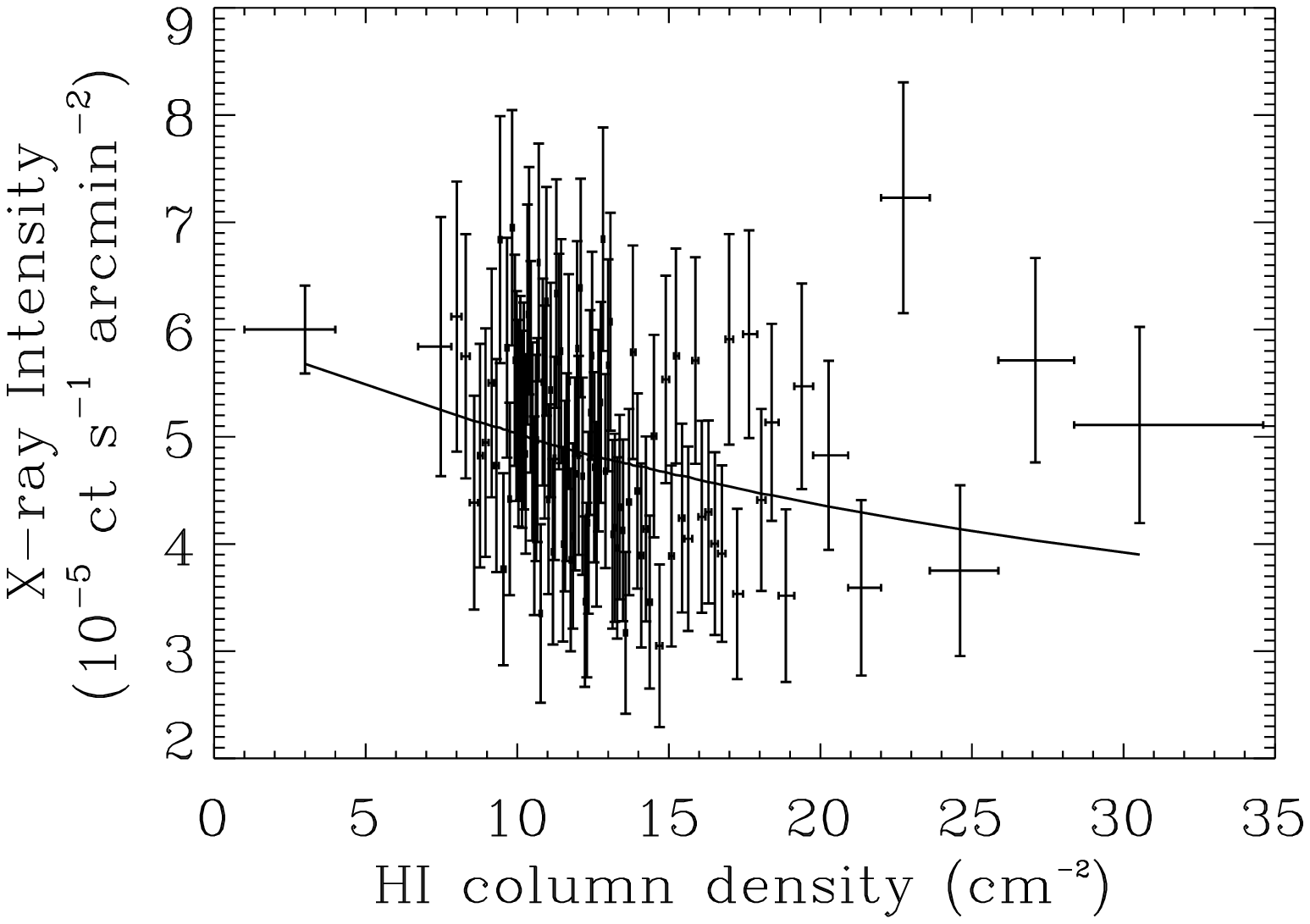}
\caption{(URL: 
http://www.astro.nwu.edu/astro/wqd/bridge/fig4.gif) 
0.7~keV-band X-ray intensity vs. atomic hydrogen column density of the cloud. 
The Galactic neutral gas component is not plotted, but is included in
the fit. The horizontal bars show atomic hydrogen column density intervals
of the bins, and the vertical bars are $1\sigma$ errors.  The best-fit 
model is plotted as the solid curve.
\label{fig4}}
\end{figure}

	As expected, we find no significant anti-correlation between
\ion{H}{1} and X-ray intensity, except in the R4 and R5 bands. 
The detection of the extragalactic background in the 
R5 band is marginal at $\sim 95\%$ confidence; the best-fit intensity is
$2.4 (0.-5.0)\times 10^{-5} {\rm~counts~s^{-1}~arcmin^{-2}}$ (90\% 
confidence limits).

	We thus focus on results obtained in the R4 band.
The best-fit ($\chi^2/d.o.f = 96.5/99$; Fig.~4) gives
$I_g$ and $I_e$ as $2.7 (1.5-3.7) \times 10^{-5} $ and 
$4.0 (2.0-6.2) \times 10^{-5} {\rm~counts~s^{-1}~arcmin^{-2}}$;
the 10\% residual source flux has been subtracted from
$I_e$. Assuming again the power law 
($\alpha = 1$), the corresponding extragalactic energy intensity is
$20 (10-31) {\rm~keV~s^{-1}~cm^{-2}~keV^{-1}} {\rm~sr^{-1}}$ at 0.7~keV. 
To obtain an estimate of the total extragalactic background,
one should include $7.5 {\rm~keV~s^{-1}~cm^{-2}~keV^{-1}} {\rm~sr^{-1}}$,
an expected contribution from sources above our source subtraction threshold
of $7 \times 10^{-15} {\rm~ergs~s^{-1}~cm^{-2}}$ (H93). 

	Although the fit in Fig.~4 is satisfactory, systematic
effects still need some consideration. The most important one is probably 
associated with the discrete 
source contribution. Our imperfect source subtraction is at least
partially responsible for apparent outliers seen in Fig.~4.
For example, the bin at $ N_{HI} \sim 2.3 \times 10^{21} {\rm~cm^{-2}}$
deviates by $2.8\sigma$ from the best-fit model,
and appears to be influenced by the presence of two peaks with 
signal-to-noise ratios $ \sim 2$ in Fig.~3. 
The removal of this bin would slightly increase our estimate of the 
extragalactic background. In addition, the extragalactic background 
may contain structures on scales comparable to the size
of the cloud (e.g., Burg et al 1992). As a test, we have analyzed the 
\rosat deep survey observation in the Lockman Hole (wp900029; H93), 
following the same procedure as 
described above. Pretending that the observation is our
on-cloud observation, we detected no significant anti-correlation between 
X-ray and \ion{H}{1}. Nevertheless, the issue
regarding the possible structure of the extragalactic background
remains to be explored, but is beyond the scope of this paper. 
For the time being, we assume that these systematic effects are 
insignificant.

\section {Discussion}
	
	The above results represent the first separation of the Galactic 
foreground from the extragalactic background at $\sim 0.7$~keV. The
Galactic foreground measurement places a fundamental limit on the integrated
emission in the Galaxy, including its halo. In the following, however, we  
concentrate on discussing the implication of the extragalactic background
measurement on the hot IGM.

	Is there any evidence for the presence of a {\sl diffuse}
0.7~keV extragalactic background? To the lowest source detection
limit of $\sim 2.5 \times 10^{-15} {\rm~ergs~s^{-1}~cm^{-2}}$ (0.5-2~keV), 
 the integrated mean source contribution is 
8.8 ${\rm~keV~s^{-1}~cm^{-2}~keV^{-1}} {\rm~sr^{-1}}$ at 0.7~keV, only
15\% of which is below our source removal threshold. 
This contribution is significantly below our 95\% confidence lower limit to the
total extragalactic background 18 ${\rm~keV~s^{-1}~cm^{-2}~keV^{-1}} 
{\rm~sr^{-1}}$. Further extrapolating the Hasinger et al. source flux 
distribution (H93) to 
$\sim 5 \times 10^{-18} {\rm~ergs~s^{-1}~cm^{-2}}$, a limit at which
all X-rays observed in the 1-2 band are attributed to point-like sources
(H93), the source contribution would then be $14 
{\rm~keV~s^{-1}~cm^{-2}~keV^{-1}} {\rm~sr^{-1}}$, 
still considerably below our limit. Because the mean source spectrum 
becomes harder (i.e., $\alpha \lesssim 1$) with decreasing flux (H93; 
Wang \& McCray 1994; Vikhlinin et al. 1995), the real source contribution
should be lower. 
Therefore, a diffuse contribution $\gtrsim 4 {\rm~keV~s^{-1}~cm^{-2}~keV^{-1}}
{\rm~sr^{-1}}$ is needed to explain our measured 0.7~keV extragalactic 
background. 

	Evidence for a {\sl thermal} component of the 0.7~keV background comes 
from the {\sl ASCA} X-ray Observatory, which provides spectroscopic 
capability between 0.4-10~keV. The spectrum
of the X-ray background clearly shows an excess below 1~keV 
above the extrapolation of the single power law ($\alpha \approx 0.4$)
that fits well to the 1-10~keV spectrum, and also
shows evidence of \ion{O}{7} and \ion{O}{8}
lines, suggesting that at least part of the excess is thermal 
in origin (Gendreau et al. 1995). There are, however, some significant
discrepancies ($\sim 30\%$) between soft X-ray background normalizations 
derived from different instruments (McCammon \& Sanders 1990; Wu et al. 1991;
Garmire et al. 1992; Gendreau et al. 1995; Snowden et al. 1995),
making it hard to compare different measurements of absolute background 
intensities. Nevertheless, our \rosat measurement of the extragalactic 
background, combined with the \asca spectral results, do seem to 
suggest that a considerable fraction of the 0.7~keV background
arises in extragalactic diffuse thermal gas.

	While the exact thermal state of the gas is yet to be determined,
we here characterize the gas with an
optically-thin thermal plasma (Raymond \& Smith 1977). The mean metallicity 
of the gas should fall within the range between 
$\sim 35\%$ solar for rich clusters 
and $\sim 3.5\%$ for voids (CKOR95 and references therein).
We choose 6\% solar as a representative value. Assuming a temperature of
$10^{6.8}$~K, at which the gas emission peaks in the R4-band, we obtain an 
emission measure of $\sim 7 \times 10^{-3} {\rm~cm^{-6}~pc}$ from
our best estimate $14 {\rm~keV~s^{-1}~cm^{-2}~keV^{-1}} 
{\rm~sr^{-1}}$ of the diffuse extragalactic 0.7~keV background.
For gas with a distribution of temperatures, the gas contribution to the  
background would be less efficient and thus the inferred emission measure 
would be higher. 

	Where is the diffuse gas located? The closest possibility is  
the intragroup space of our Local Group of galaxies. Could the gas 
be the hot intragroup medium? Apparent diffuse X-ray
emission has been detected in a number of groups. But all these groups 
contain large elliptical galaxies with $L_B \gtrsim 5 \times 10^{10} 
L_\odot$ (Mulchaey et al. 1996). The absence of such a galaxy in our Local 
Group suggests that the Local Group does not contribute significantly to 
the X-ray background. Therefore, the diffuse gas is likely cosmological 
in origin.

	We speculate that the 0.7~keV diffuse extragalactic background
may represent the integrated emission from the diffuse IGM. 
Away from those more-or-less virialized systems, such as rich clusters of
galaxies, the diffuse IGM is believed to be the major component of the 
large-scale structure, and may
account for most of the baryonic content of the universe required by 
Big Bang nucleosynthesis (e.g., Copi, Schramm, \& Turner 1995; 
Tytler, Fan, \& Burles 1996); the total visible mass in galaxies and 
in the hot intracluster medium is $\lesssim 10\%$ of the content (e.g.,
Persic \& Salucci 1992).  Detailed hydrodynamic simulations (e.g., OC96)
show that a substantial fraction of the IGM should be within a temperature 
range of $10^6-10^7$~K. This hot IGM may contribute 
significantly more than individual rich clusters to the background at 
$\sim 0.7$~keV, where various spectral lines are present
(mainly \ion{O}{7}, \ion{O}{8}, and the iron blade;
CKOR95). Our estimate of the diffuse 
extragalactic emission favors models that predict relatively large
contributions ($\gtrsim 15\%$) from the IGM 
to the 0.7~keV background (e.g., the COBE-normalized, flat 
CDM model with a cosmological constant; CKOR95). But our
result alone does not provide tight constraints on model parameters. So let us
also check other available limits on the extragalactic soft X-ray background.

Many studies of X-ray shadows have already been carried out in the PSPC R1+R2 
band, which is sensitive to photons $\sim 0.25$~keV (e.g., Burrows \& 
Mendenhall 1991; Snowden et al 1994b; Wang \& Yu 1995; 
Cui et al. 1996). In particular, based on X-ray shadows cast by parts of 
the nearby galaxy NGC 3184 and by high latitude clouds in Ursa Major
(see also Snowden et al 1994b), Cui et al. have derived lower and upper 
limits to the total extragalactic background as
32 and 65 ${\rm~keV~s^{-1}~cm^{-2}~keV^{-1}} {\rm~sr^{-1}}$ at 0.25~keV. 
The lower limit is essentially the same as the integrated contribution 
from directly detected sources to the flux limit of 
$2.5 \times 10^{-15} {\rm~ergs~s^{-1}~cm^{-2}}$ (H93). The upper limit
 allows a substantial diffuse contribution.
With the power law of $\alpha =1$ for the mean source spectrum, 
the total source contribution that accounts for all intensity
in the 1-2~keV band corresponds to $\sim 42 {\rm~keV~s^{-1}~cm^{-2}~keV^{-1}} 
{\rm~sr^{-1}}$ at 0.25~keV. With a flatter spectrum, the contribution
would be lower. But, Galactic hot gas beyond the 
high-latitude clouds may also contribute. Therefore, the allowed diffuse 
extragalactic background is $\sim 23 {\rm~keV~s^{-1}~cm^{-2}~keV^{-1}} 
{\rm~sr^{-1}}$. 

	In comparison, we have here derived a diffuse 
extragalactic background intensity as 
$14 (4 - 25) {\rm~keV~s^{-1}~cm^{-2}~keV^{-1}} {\rm~sr^{-1}}$ at 0.7~keV. 
Clearly, the uncertainty in the actual intensity is substantial, 
but the 95\% confidence lower limit is conservative.
A $10^{6.8}$~K thermal plasma, accounting for the best-fit intensity, 
can contribute an intensity of $\sim 9
{\rm~keV~s^{-1}~cm^{-2}~keV^{-1}} {\rm~sr^{-1}}$ at 0.25~keV. 
Alternatively, a $10^{6.3}$~K plasma, characteristic of the diffuse
hard component (Wang \& McCray 1994), can provide $\sim 21
{\rm~keV~s^{-1}~cm^{-2}~keV^{-1}} {\rm~sr^{-1}}$, close to
the allowed diffuse background limit at 0.25~keV.  Therefore, 
a detailed comparison of the available limits with realistically simulated
IGM emission spectra may already be useful in constraining the
thermal, and possibly chemical, properties of the IGM.

\acknowledgements

	We thank Dick McCray for detailed editorial corrections and 
valuable comments, and Jianke Li for assisting in the ATCA observations.
This work is supported by a Lindheimer Fellowship and by NASA under Grants 
NAG 5-2716 and 5-2717.

\end{document}